\documentclass[
reprint,
superscriptaddress,
 amsmath,amssymb,
 prl,
]{revtex4-1}

\usepackage{graphicx}
\usepackage{dcolumn}
\usepackage{bm}
\usepackage{nicefrac}
\usepackage{caption}
\usepackage{subcaption}
\usepackage[mathlines]{lineno}

\begin{document}
\title{ Measurement Of Quasiparticle Transport In Aluminum Films Using Tungsten Transition-Edge Sensors
}

\author{J.J.~Yen}
\email[Email address: ]{jeffyen@stanford.edu}
\affiliation{Department of Physics, Stanford University, Stanford, CA 94305, USA}
\author{B.~Shank}
\affiliation{Department of Physics, Stanford University, Stanford, CA 94305, USA}
\author{B.A.~Young}
\affiliation{Department of Physics, Santa Clara University, CA 95053, USA}
\author{B.~Cabrera}
\affiliation{Department of Physics, Stanford University, Stanford, CA 94305, USA}
\author{P.L.~Brink}
\affiliation{SLAC National Accelerator Laboratory, 2575 Sand Hill Road, Menlo Park, CA 94025, USA}
\author{M.~Cherry}
\affiliation{SLAC National Accelerator Laboratory, 2575 Sand Hill Road, Menlo Park, CA 94025, USA}
\author{J.M.~Kreikebaum}
\affiliation{Department of Physics, Stanford University, Stanford, CA 94305, USA}
\affiliation {Department of Physics, Santa Clara University, CA 95053, USA}
\author{R.~Moffatt}
\affiliation{Department of Physics, Stanford University, Stanford, CA 94305, USA}
\author{P.~Redl}
\affiliation{Department of Physics, Stanford University, Stanford, CA 94305, USA}
\author{A.~Tomada}
\affiliation{SLAC National Accelerator Laboratory, 2575 Sand Hill Road, Menlo Park, CA 94025, USA}
\author{E.C.~Tortorici}
\affiliation{Department of Physics, Santa Clara University, CA 95053, USA}

\date{\today}

\begin{abstract}
We report new experimental studies to understand the physics of phonon sensors which utilize quasiparticle diffusion in thin aluminum films into tungsten transition-edge-sensors (TESs) operated at 35 mK. We show that basic TES physics and a simple physical model of the overlap region between the W and Al films in our devices enables us to accurately reproduce the experimentally observed pulse shapes from x-rays absorbed in the Al films. We further estimate quasiparticle loss in Al films using a simple diffusion equation approach. 
\end{abstract}


\maketitle



\section{Introduction}
Quasiparticle transport dynamics have been studied in the lab by many groups \cite{diff_study_1,diff_study_2, Trapping} using different materials, fabrication processes, and readout schemes. Quasiparticle transport in Al films plays an important role in the design specifications of Cryogenic Dark Matter Search (CDMS) detectors \cite{CDMS}. These detectors utilize photolithographically patterned films of sputtered Al and W on both sides of high-purity, kg-scale, Ge and Si crystals. The superconducting Al and W films perform two roles simultaneously: they absorb phonon energy and they serve as ionization collection electrodes. 

When a particle interacts with a CDMS detector, electron-hole pairs and phonons are created. Under typical operating conditions, a $\sim$1V/cm bias is used to drift the e-/h{\scriptsize+} pairs through the bulk of the crystal so charge can be collected at the detector surfaces. At the same time, the athermal phonons produced by the event make their way to the detector surfaces where they can be absorbed in the Al film by breaking Cooper pairs which create quasiparticles. Ideally, the quasiparticles diffuse randomly in the Al until they get trapped in the overlap region between the Al and W films, where the superconducting energy gap is smaller than in the Al film alone \cite{Booth}. This trapped energy gets absorbed by an attached W-TES, adding heat and providing the detector's phonon signal for that event. We call these phonon sensors Quasiparticle-trap-assisted-Electrothermal-feedback Transition-edge-sensors (QETs) \cite{QET}.

The quasiparticle (qp) trapping length in CDMS Al films impacts overall detector energy performance. 
Here we present results from a detailed study of energy collection and qp propagation in Al films coupled to W-TESs and describe an innovative model that explains QET pulse shapes and overall performance, and provides a way to measure qp trapping lengths in thin films and the energy transport efficiency from the qp energy to the TES electron system. Our measurements have benefited from a newly implemented signal analysis approach based on template matching rather than pulse integration which improves our energy resolution by a factor of two and yields better event reconstruction overall \cite{Ben_APL}.

\section{Experimental Setup}

\begin{figure}[h]
\begin{center}
\includegraphics[width=2.8in]{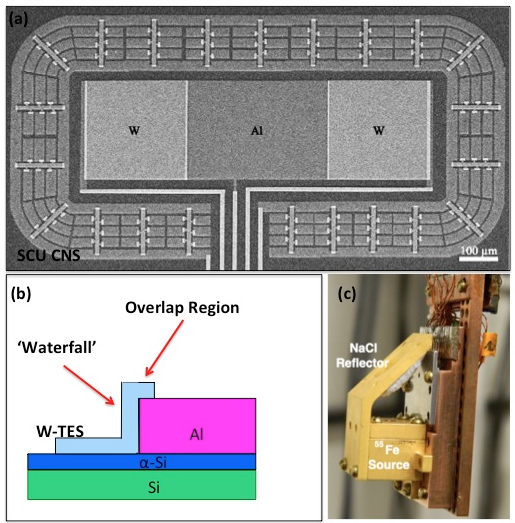} 
\end{center}
\vspace{-.2in}
\captionsetup{justification=raggedright, singlelinecheck=false}
\caption{(a) SEM image of Al/W test device. The W-TESs at the ends of the Al film are 250$\mu$m x 250$\mu$m. The racetrack-shaped outer channel acts as a veto for substrate events. (b) Schematic side view (not to scale) where each W-TES overlaps the Al film. (c) Sample mount with $^{55}$Fe / NaCl x-ray fluorescence source. The test device is hidden behind a collimator plate.}

\label{fig.fab}
\end{figure}

Test samples consisted of photolithographically patterned, 300 nm-thick Al and 40 nm-thick W films. Three Al film lengths were studied: 250$\mu$m, 350$\mu$m and 500$\mu$m. The metallization and process steps were identical to those used for CDMS detectors, including a 40 nm layer of amorphous Si (aSi) sputtered on each cleaned Si substrate just prior to metallization. 
Fig. \ref{fig.fab}a shows an image of one test device with a central 250 $\mu$m-wide x 350 $\mu$m-long Al phonon absorption film coupled to 250 $\mu$m x 250 $\mu$m W TESs (W-TES1 and W-TES2) at either end. A distributed racetrack-like outer TES channel (W-TES3) served as a veto for substrate events. A schematic diagram of the film geometry at the overlap regions between the W-TESs and the Al energy collection film is shown in Fig. \ref{fig.fab}b. Fig. \ref{fig.fab}c shows the OFHC Cu structure used to both anchor devices to the mixing chamber of our dilution refrigerator and expose a single device (through collimators) to an $^{55}$Fe/NaCl fluorescence source (Cl K$\alpha$ at 2.62 keV). With this arrangement, low energy source x-rays reached our devices $\sim$ 20 times per second. 


\section{W-TES Energy Collection}
Collimated x-ray absorption events were measured using a conventional voltage-biased TES circuitry setup \cite{QET}, with the W-TES sensor biased in the steepest part of its resistive transition. The total change in internal energy of a TES under such conditions is well approximated by:
\begin{equation}
\Delta U = \Delta U_{ext} + \Delta U_{Joule} + \Delta U_{e-ph} = 0,
\end{equation}
where $\Delta U_{ext}$ represents the deposited x-ray energy, $\Delta U_{Joule}$ corresponds to the Joule heating $\sim$ {V$^2$/R} of the biased TES, and $\Delta U_{e-ph}$ is an energy loss term arising from electron-phonon coupling within the TES. This latter term accounts for the thermal relaxation of the TES. It is relatively small when the TES is operated in the linear, non-saturated region of its $R(T,I)$ curve and small energy inputs are considered. In general, event energy absorbed by a voltage-biased TES will increase sensor resistance and thus decrease the instantaneous energy loss from Joule heating. When in the linear, low energy regime, the first two terms in the energy balance equation dominate the physics, and essentially cancel each other. However, when the energy flux into a TES is sufficient to drive the TES fully normal, $\Delta U_{e-ph}$ can be significant. Below, we show that by consistently including the $\Delta U_{e-ph}$ term in our model we can more accurately reproduce the observed pulse shapes and energy distributions of W-TES events in both the non-saturated and saturated regimes \cite{Ben_APL}.

\begin{figure}[h]
 \begin{center}
 \includegraphics[width=3.5in]{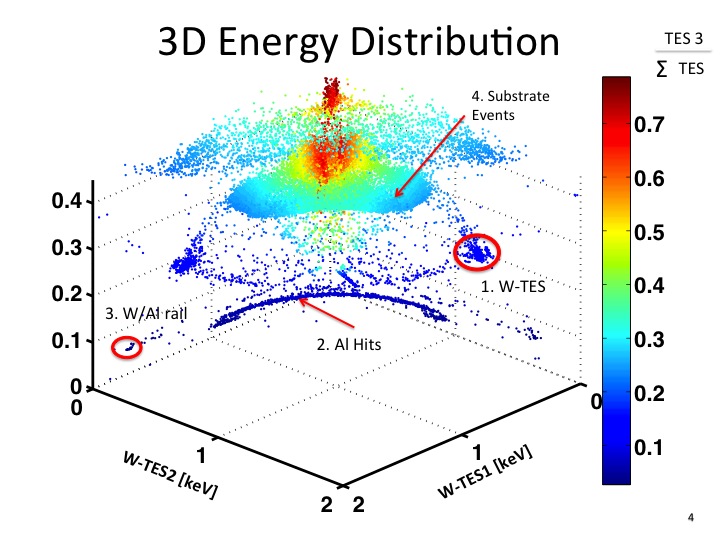} 
 \vspace{-.3in}
 \end{center}
 \captionsetup{justification=RaggedRight, singlelinecheck=false}
 \caption{X-ray event energy collected in each of the three W-TESs of a 250 $\mu m$-long central Al film device. Four distinct x-ray interaction locations are noted: W-TES, central Al, Al/W overlap regions, and the substrate. The color bar indicates the fraction of the total detected energy appearing in the substrate channel (W-TES3). The energy collected by W-TES1 and W-TES2 for x-ray hits along the central Al film (the banana-shaped cluster of points shown) is consistent with the known device geometry.}
 \label{fig.3d_banana}
\end{figure}

Fig. \ref{fig.3d_banana} shows the energy detected by each of the three W-TESs on a single test device exposed for $\sim$48 hours to our NaCl fluorescence source using the set-up shown in Fig. \ref{fig.fab}c. The data were obtained with a 250 $\mu m$-long Al film device similar to that shown in Fig. \ref{fig.fab}a. Event energies were determined using a non-linear optimal filter template fitting approach \cite{Ben_APL}. As shown in Fig. \ref{fig.3d_banana}, we observed four basic classes of events: (1) x-rays absorbed directly in W-TES1 or W-TES2, (2) x-rays absorbed in the central Al film, (3) x-rays absorbed in one of the four main W/Al overlap regions of the device (one at each end of both W-TES1 and W-TES2), and most commonly (4) x-rays absorbed in the Si substrate (large W-TES3 signal). The relative count rates observed for the various event types were consistent with the source-collimator geometry and the known penetration depths \cite{Mass_Attenuation} for 2.62 keV x-rays in Al (3.3 $\mu m$) and W (0.2 $\mu m$).

We scaled event energy measurements to the initial energy stored in qps only after their number became constant, {\it i.e.} after the initial fast phonon decay modes were complete but before qps shed sub-gap phonons \cite{Goldie}.
In our experiments, a maximum of only 1.42 keV of the incident 2.62 keV Cl K$\alpha$ x-ray energy was collected in W-TES1, even for a direct-hit in that sensor (see Fig. \ref{fig.3d_banana}). 
This large energy deficit can be explained using an energy down-conversion model recently published by Kozorezov,{\it et.al.} \cite{phonon_loss}. Their model defines three stages of the energy down-conversion process following the absorption of an x-ray in a thin metal film. 
The most relevant to our experiments with W-TESs is Stage II, where athermal phonon leakage into the substrate dominates the film's energy loss to the substrate. Stage II can be subdivided into two main parts. In the first part, the mean energy of electronic excitations, $\epsilon$, is below some threshold, $E_1^*$, but much higher than the Debye energy: $\Omega_D<<\epsilon< E_1^*$. In this regime, energy loss to the substrate can be strongly dependent on event location in the film ({\it i.e.} proximity to the film-substrate boundary) and spectral peaks get broadened, but not typically shifted appreciably in energy.

The second part of Stage II is characterized by $\Omega_D > \epsilon > \Omega_1$, where $\Omega_1$ is a low-energy threshold above which electron and hole relaxation by phonon emission is still important, but below which the dynamics is again dominated by electronic interactions. This portion of the energy cascade process turns out to be more important than expected for explaining the observed energy loss in TESs and other film-based devices. Applying Eqs. 7, 9 and 10 of Ref. \cite{phonon_loss} to our experimental conditions yields a predicted fractional energy loss in our W films of 49\% for direct-hit x-ray events. In our experiments we observe an actual energy loss of $\sim$ 43\% for these direct-hit events. One effect that can reduce this small discrepancy is the reabsorption of high-energy escape phonons back into the W-TES from the substrate. 

\section{Pulse Shapes: Waterfall model}
\begin{figure}[h]
  \begin{center}
  \includegraphics[width=3.5in]{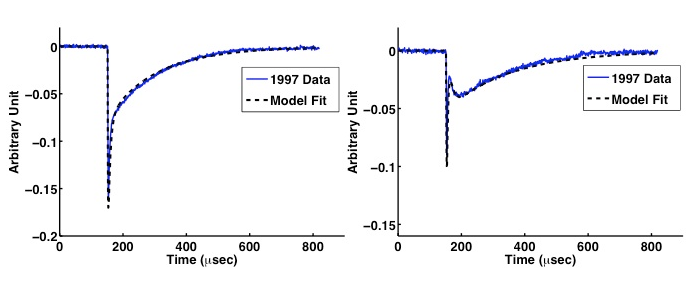} 
  \vspace{-.3in}
  \end{center}
  \captionsetup{justification=raggedright, singlelinecheck=false}
  \caption{Overlay of raw data and simulated pulses for: (a) a typical Al/W test device, (b) a similar Al/W device, first tested in 1997, with odd pulse shapes that we now understand.}
  \label{fig.Pulses}
\end{figure}

We have developed a simple physical model that accurately describes the pulse shapes observed with our Al/W devices. We show in Fig. \ref{fig.Pulses}a one simulated pulse from this model superimposed on a raw pulse from a well-behaved device like the one shown in Fig. \ref{fig.fab}a. We have also used this model to reproduce previously unexplained pulse shapes \cite{LTD11} obtained with a device of similar design that was studied first in 1997 and then again in 2014. The same, unusual pulse shapes were observed in both data sets. The remarkable double-peak structure for that device is shown in Fig. \ref{fig.Pulses}b. 

\begin{figure}[h]
  \begin{center}
  \includegraphics[width=3.0in]{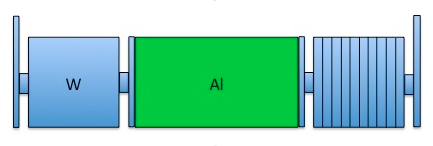} 
  \vspace{-0.3in}
  \end{center}
  \captionsetup{justification=raggedright, singlelinecheck=false}
  \caption{Physical model of our Al/W device that: (a) models imperfect interfaces between the Al and W regions as resistive 'bottlenecks' that can affect the critical current and the TES response function, and (b) treats W-TES1 and W-TES2 each as a series of ten parallel strips with thermal conductance between the strips given approximately by the Wiedemann-Franz Law.}
  \label{fig.waterfall_Betty}
\end{figure}

The key elements of our physical model are shown in Fig. \ref{fig.waterfall_Betty}. In the model, weak links of W are used to mimic the step-coverage impedance where the 40 nm-thick W film overlaps the 300 nm-thick Al film below it, as the W transitions down to the substrate where it operates as a TES (see Fig. \ref{fig.fab}b). We refer to these film transition regions as ``waterfall'' regions based on their appearance in SEM images \cite{LTD15_Jeff}. In our test devices, the W/Al overlap region is typically excellent along the top surface of the Al but more limited on the steep Al sidewalls. Our model treats the added impedance of an imperfect waterfall region as a weak W link that acts effectively as a small Joule heater providing constant power even when the W-TES itself is in its superconducting transition. This impedance alters the superconducting temperature and critical current of the TES in predictable ways. 
Additionally, in our model each TES square is divided into ten equal-width strips parallel to the W/Al overlap region. The Wiedemann-Franz Law is then used in a one-dimensional (1D) simulation of qp thermalization in the voltage-biased TES as energy flows through it laterally. 

Our new model works well. For example, it yields the first decay-time in the raw data pulse shown in Fig. \ref{fig.Pulses}a. It also correctly predicts the second distinct decay-time that corresponds to the time ($\tau_{etf}$) needed for the TES to cool back to its equilibrium state. Lastly, the model explains the double-peaked pulses observed with our older devices from 1997 - the odd pulse shapes we now know resulted from poor film connectivity between each W-TES and its corresponding Al bias line at the end away from the main Al absorber (see Fig. \ref{fig.waterfall_Betty}). A detailed description of this new model and its successful use in pulse shape simulations is discussed in Ref. \cite{Ben_APL}.

\section{Quasiparticle transport: Diffusion, Absorption and Energy Collection}
\begin{figure}[h]
  \begin{center}
  \includegraphics[width=2.8in]{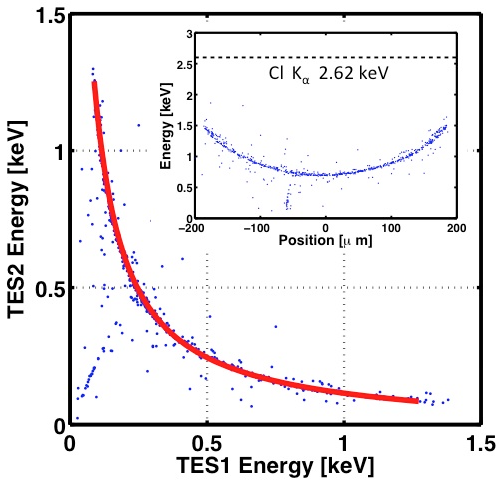} 
  \vspace{-.2in}
  \end{center}
  \captionsetup{justification=raggedright, singlelinecheck=false}
  \caption{Overlay of raw energy collection distribution and Maximum Likelihood fit. The banana-shaped cluster of points corresponds to direct-hit x-rays in the main Al film. (Inset): Collected x-ray energy vs. event location along the Al film. The cluster of points near -55$\mu$m is consistent with x-rays absorbed in the ground line of the main Al film.}
  \label{fig.banana_fit}
\end{figure}
After selecting Al direct-hit events (dark blue in Fig. \ref{fig.3d_banana}) using the method described in Ref. \cite{LTD15_Jeff}, we modeled qp transport in the Al film using a 1D diffusion equation with a linear loss term:
\begin{equation}
\frac{\partial{n}}{\partial{t}} = D_{\textrm{Al}} \space \frac{\partial^2{n}}{\partial{x^2}} - \frac{n}{\tau_{\textrm{Al}}} + s,
\label{diffusion_eqn}
\end{equation}
where $n = n(x,t)$ is the linear number density of qps, $D_{\textrm{Al}}$ is the diffusivity of qps, and $\tau_{\textrm{Al}}$ is the qp trapping time. The source term $s= q\, \delta(x-x_0) \delta(t-t_0)$ represents the rate of qp density creation. The rates for qp absorption into W-TES1 and W-TES2, symbolized by $I_1$ and $I_2$ respectively, were modeled by the linear relations:
\begin{equation}
I_1 = n_{1}\,v_{\textrm{1}}, \ \ I_2 = n_{2}\,v_{\textrm{2}},
\label{absorption_eqn}
\end{equation}
where the coefficient $v_{\textrm{1}}(v_{\textrm{2}}) $ has units of length/time, and $n_{1}$ ($n_{2}$) is the qp number density at the W/Al boundary closest to W-TES1 (W-TES2). This 1D approach is sufficient because the qps are reflected at the edges of the Al, and the mean free path is smaller than the width of the film, making diffusion along the two axes independent. 

Equation \ref{diffusion_eqn} can be solved analytically to find the fraction F$_{1}$(F$_{2}$) of qp generated by an event that is absorbed in W-TES1(W-TES2):

\footnotesize
\begin{eqnarray}
F_1 &=&  \frac{\Lambda_d\left( \lambda_{2}\, \textrm{cosh}\left(\frac{1+2\xi}{2\Lambda_d} \right) + \Lambda_d\, \textrm{sinh}\left( \frac{1+2\xi}{2\Lambda_d} \right) \right)}{\Lambda_d(\lambda_{1} + \lambda_{2})\, \textrm{cosh} \left( \frac{1}{\Lambda_d}\right) + (\Lambda_d^2 + \lambda_{1}\lambda_{2})\,\sinh \left(\frac{1}{\Lambda_d} \right) }
\label{fraction1}
\\
F_2 &=& \frac{\Lambda_\textrm{d}\left( \lambda_{1}\, \textrm{cosh}\left(\frac{1-2\xi}{2\Lambda_d} \right) + \Lambda_d\, \textrm{sinh}\left( \frac{1-2\xi}{2\Lambda_d} \right) \right)}{\Lambda_d(\lambda_{1} + \lambda_{2})\, \textrm{cosh} \left( \frac{1}{\Lambda_d}\right) + (\Lambda_d^2 + \lambda_{1}\lambda_{2})\,\textrm{sinh}\left(\frac{1}{\Lambda_d} \right) }
\label{fraction2}
\end{eqnarray}
\normalsize
The dimensionless variable $\Lambda_\textrm{d} \equiv L_d / L$ depends on the characteristic diffusion length $L_d = \sqrt{D_\textrm{Al}\tau_\textrm{Al}}$ of the Al film, and the term $\xi \equiv x_0/L$ depends on the qp source location, $x_0$, measured from the center of the Al film. $L$ is the length of the Al film.
The dimensionless parameters $\lambda_1$ and $\lambda_2$ are defined by the relation, $ \lambda_{i} \equiv L_{i}/ L$, where L$_i = D_\textrm{Al}/v_i$ $ (i =1,2 )$ is a characteristic qp absorption parameter with units of length that varies inversely with the efficiency for coupling qp into each W-TES. In general the W-TESs would have slightly different qp absorption capabilities, hence $\lambda_{1} \neq \lambda_{2}$. However, if one assumes the same absorption capability for the two TESs, Eq. \ref{fraction1} and Eq. \ref{fraction2} can be further simplified to the form shown in Eq. 1 of Ref. \cite{diff_study_1}. 

Fig. \ref{fig.banana_fit} shows a Maximum Likelihood fit of this diffusion model to x-ray data for a 350 $\mu$m-long Al film. The fit yields estimates for three important parameters: the characteristic qp diffusion length, $L_d$, the qp absorption into W-TESs, $L_{\textrm{1}}(L_\textrm{2})$, and an energy scaling factor, $\mathcal{E}_{\textrm{sf}}$. The scaling factor corresponds to the deposited energy before position dependent qp trapping and sub-gap phonon losses have occurred as energy is absorbed into the two W-TESs.
Applying Eq. \ref{diffusion_eqn} to our data yields $L_{\textrm{d}} \sim 130 \mu$m for three Al film lengths studied: 250 $\mu$m, 350 $\mu$m, and 500 $\mu$m. For small values of $L_i$, the band of Al direct-hit events shown in Fig. \ref{fig.banana_fit} would extend towards the energy axes. In our data, $L_\textrm{1}$ $\approx$ $L_\textrm{2}$ $\sim$100 $\mu$m, and we observe gaps between the end points of the Al direct-hit band and the energy axes. Summing the two W-TES energies and reconstructing position yields the inset of Fig. \ref{fig.banana_fit}. 

\begin{figure}[h]
  \begin{center}
  \includegraphics[width=3.in]{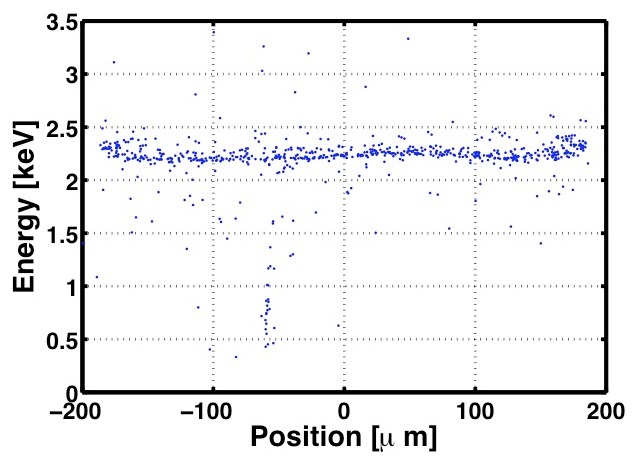} 
  \vspace{-.2in}
  \end{center}
  \captionsetup{justification=raggedright, singlelinecheck=false}
  \caption{Reconstructed Cl K$_\alpha$ x-ray energy as a function of event position along Al film. This corresponds to the deposited energy before position dependent qp trapping and sub-gap phonon losses have occurred.  }
  \label{fig.position}
\end{figure}

Fig. \ref{fig.position} shows the reconstructed energy vs. position data of Fig. \ref{fig.banana_fit} using the parameters from our diffusion model fit. The scaling factor obtained from the model yields a total event energy of 2.3 keV rather than the expected 2.62 keV. This $\sim $ 10$\%$ discrepancy is consistent with known energy down-conversion mechanisms. The 5\% variation in reconstructed energies shown in Fig. \ref{fig.position} can be corrected using a model that includes the latter stages ($\epsilon < 3 \Delta$) of the energy down-conversion cascade and simulates qp trapping in terms of a percolation threshold (below which qps are trapped by local variations in the gap) \cite{percolation}.

\section{Conclusion}
Our new TES model accurately estimates the energy of direct-hit x-rays in W-TESs. The results are consistent with phonon and qp energy down-conversion physics and the model also provides a better understanding of the processes needed to improve the energy transport in CDMS Al films. In the simple diffusion model used in this work, losses to sub-gap phonons and qp trapping were combined into a single, generic term. A more detailed study that includes percolation threshold effects from spatial variations in the superconducting gap of our Al films will be reported soon. The 0.3 $\mu$m-thick Al films used in this study have a measured qp characteristic diffusion length $L_{\textrm{d}} \sim$130$\mu$m which is thickness-limited. We are presently using SEM and FIB imaging tools to appropriately modify detector fabrication recipes in order to improve connectivity at the Al/W interfaces, allowing detectors to be made in the future with Al films that are twice the current thickness.

\section{Acknowledgements}
\begin{acknowledgements}
We are grateful to the Stanford Physics Machine Shop for making source and sample holders, and collimators. We thank K. D. Irwin and S. Chaudhuri for useful discussions on TES physics. We also thank M. Pyle and K. Schneck for CDMS related conversations. The authors would also like to thank A. Kozorezov and S. Bandler for useful qp and phonon physics discussions. 
We acknowledge support from the Department of Energy, DOE grant DE-FG02-13ER41918, and the National Science Foundation, NSF grant PHY-1102842.
\end{acknowledgements}


\begin{thebibliography}{99}

\bibitem{diff_study_1}
 M. Loidl, S. Cooper, O. Meier, F. Pr\"{o}bst, G. S\'{a}fr\'{a}n, W. Seidel, M. Sisti, L. Stodolsky, S. Uchaikin, {\it Nucl. Instr. and Meth. A,} \textbf{465}, 440-446, (2001).

\bibitem{diff_study_2}
J. Martin, S. Lemkea, R. Grossa, R.P. Huebenera, P. Videlerb, N. Randob, T. Peacockb, P. Verhoeveb, F.A. Jansenb, {\it Nucl. Instrum. Methods A} \textbf{370}, 88-90, (1996).

\bibitem{Trapping}
C. Bailey, J. Adams, S. Bandler, J. Chervenak, M. Eckart, A. Ewin, F. Finkbeiner, R. Kelley, C. Kilbourne, F. Porter, J. Sadleir, S. Smith, M. Sultana, {\it J. Low Temp.} \textbf{167}, 3-4, (2012).



\bibitem{CDMS}
Z. Ahmed et al., {\it Phys. Rev. Lett. } \textbf{106}, 131302, (2011).

\bibitem{Booth}
N. E. Booth, {\it Appl. Phys. Lett} \textbf{50}, 293, (1987).

\bibitem{QET}
K. D. Irwin, S. W. Nam, B. Cabrera, B. Chugg and B. A. Young, {\it Rev. Sci. Instrum.} \textbf{66}, 5322, (1995).

\bibitem{Ben_APL}
B. Shank, et. al., "Nonlinear Optimal Filter Technique for Analyzing Energy Depositions in TES Sensors Driven Into Saturation", this publication

\bibitem{Mass_Attenuation}
S. M. Seltzer, {\it Radiation Research.} 136-147 (1993).

\bibitem{Goldie}
Guruswamy, D J Goldie and S Withington, Supercond. Sci. Technol. 27, 055012 (2014).

\bibitem{phonon_loss}
A. G. Kozorezov, C. J. Lambert, S. R. Bandler, M. A. Balvin, S. E. Busch, P. N. Nagler, J. P. Porst, S. J. Smith, T. R. Stevenson, and J. E. Sadleir, {\it Phys. Rev. B} \textbf{87}, 104504, (2013).


\bibitem{LTD11}
M. Pyle, P. L. Brink, B. Cabrera, J. P. Castle, P. Colling, C. L. Chang, J. Cooley, T. Lipus, R. W. Ogburn, B. A. Young, {\it Nucl. Instrum. Methods A} \textbf{559}, 405-407, (2006).


\bibitem{LTD15_Jeff}
J. J. Yen, B. A. Young, B. Cabrera, P. L. Brink, M. Cherry, R. Moffatt, M. Pyle, P. Redl, A. Tomada and E. C. Tortorici {\it Journal of Low Temp. Phys.} \textbf{176}, 168-175, (2014).

\bibitem{percolation}
B. Cabrera, {\it et. al.}, in preparation.

\end{thebibliography}
\end{document}